\documentclass[twocolumn,aps,prl,showpacs,preprintnumbers,amsmath,amssymb,10pt]{revtex4-1}

\usepackage{graphicx}
\usepackage{dcolumn}
\usepackage{setspace}
\usepackage{bm}
\usepackage{multirow}

\begin{document}

\title{Enhanced dynamics in fusion of neutron-rich oxygen nuclei at above-barrier energies}

\author{S. Hudan}
\author{R.~T. deSouza}
\email{desouza@indiana.edu}
\affiliation{%
Department of Chemistry and Center for Exploration of Energy and Matter, Indiana University\\
2401 Milo B. Sampson Lane, Bloomington, Indiana 47408, USA}%

\author{A.~S. Umar}
\affiliation{%
Department of Physics and Astronomy, Vanderbilt University\\
Nashville, Tennessee 37235 USA}%

\author{Z. Lin}
\affiliation{%
Department of Physics and Center for Exploration of Energy and Matter, Indiana University\\
2401 Milo B. Sampson Lane, Bloomington, Indiana 47408 USA}%
{\affiliation{%
Department of Physics, Arizona State University, \\ 450 E. Tyler Mall, Tempe, AZ 85287-1504 USA}%

\author{C.~J. Horowitz}
\affiliation{%
Department of Physics and Center for Exploration of Energy and Matter, Indiana University\\
2401 Milo B. Sampson Lane, Bloomington, Indiana 47408, USA}%

\date{\today}

\begin{abstract}
Above-barrier fusion cross-sections for an isotopic chain of oxygen isotopes with A=16-19 incident on 
a $^{12}$C target are presented. Experimental data are compared with both static and dynamical microscopic 
calculations. These calculations 
are unable to explain the $\sim$37\% increase in the average above-barrier fusion cross-section 
observed for $^{19}$O as compared to $\beta$-stable oxygen isotopes.  
This result suggests that for neutron-rich nuclei existing time-dependent Hartree-Fock calculations underpredict the role of 
dynamics at near-barrier energies. High-quality
measurement of above-barrier fusion for an isotopic chain of increasingly neutron-rich nuclei provides an effective means to probe this fusion dynamics.
\end{abstract}


\maketitle
Nuclei with an exotic neutron-to-proton ratio provide a unique opportunity to test our microscopic 
understanding of nuclei.
A distinguishing characteristic of these nuclei is the difference in the spatial extent of 
their neutron and proton density distributions. By systematically comparing nuclei along an isotopic chain one can gain
insight into the evolution of nuclear properties with neutron number. This approach resulted in the discovery of
the halo nature of $^{11}$Li \cite{Tanihata85b}.
To date, considerable effort has been expended to understand 
ground-state properties e.g. size, shape, two neutron separation energies, pairing correlations etc. 
of these exotic nuclei \cite{Tanihata85a,Estrade14,Bagchi19,Blank93}.
As neutron-rich nuclei play a key role in r-process nucleosynthesis reactions \cite{kajino2019}, it is
crucial to understand not only their ground state properties but the response of these weakly bound systems to perturbation.
Fusion reactions provide a powerful means to assess the response of neutron-rich nuclei to perturbation. 
As fusion involves the interplay of the repulsive Coulomb and attractive nuclear potentials, by 
examining fusion for an isotopic chain one probes the neutron density distribution and 
how that density distribution evolves as the two nuclei approach and overlap \cite{Singh17,Vadas18}.
Fusion of neutron-rich nuclei has also been hypothesized 
as providing a heat source that triggers an X-ray superburst
in an accreting neutron 
star \cite{Horowitz08}. Investigating the fusion of neutron-rich nuclei is thus also important for understanding such 
astrophsyical events.
Radioactive beam facilities, both those in 
existence \cite{RIKEN,GANIL,NSCL} as well as those on the horizon \cite{FRIB} make such measurements feasible 
for the first time.

In the present work we examine 
the dependence of the above-barrier fusion cross-section 
for $^{16,17,18,19}$O ions on a $^{12}$C target. As only above-barrier energies
will be experimentally accessible for the most neutron-rich beams, it is 
important to assess how sensitively changes in neutron number impact the 
fusion cross-section in this energy regime. 
This above-barrier cross-section
can be related to 
an interaction radius at which the projectile and target nuclei fuse i.e. 
the dynamical size of the system as it fuses. 
Comparison of fusion at near and sub-barrier energies for a wide variety of projectile/target
combinations has typically resorted to a scaling approach to disentangle static and dynamic contributions 
\cite{Canto09}. By focusing on an isotopic chain incident on a single target nucleus the need for such scaling is minimized. In this work,
the static and
dynamic contributions to the above-barrier fusion cross-section
are disentangled through direct comparison of the 
experimental results with the predictions of different microscopic models.

\begin{figure}
\includegraphics[scale=0.42]{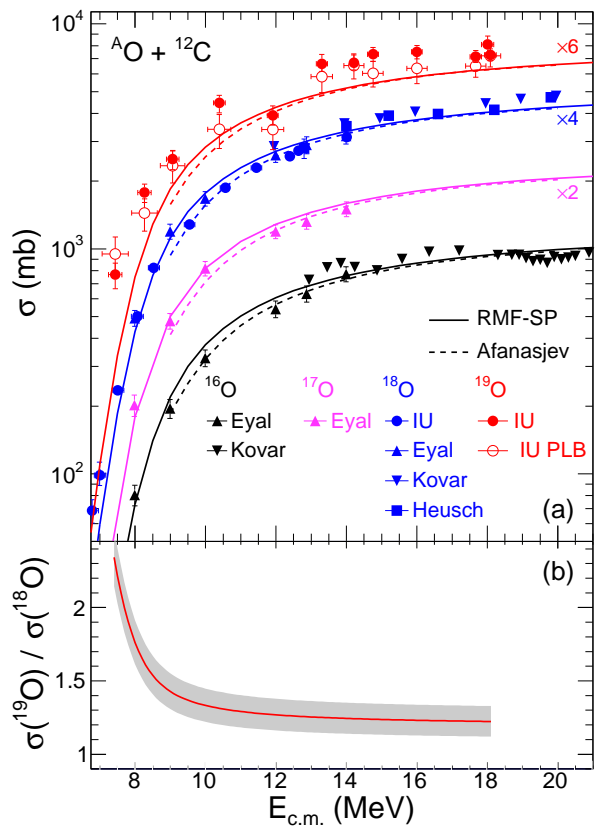}
\caption{\label{fig:fig1} 
Top panel: Fusion excitation functions for $^{16,17,18,19}$O+$^{12}$C (symbols) are 
presented along with the calculations of the RMF-SP model (solid lines).
The predictions of an analytic fusion model \cite{Afanasjev12} are indicated by the 
dashed lines.
For clarity, both data and calculations have been scaled by the 
factors shown.
Bottom panel: Ratio of the fusion cross-section $^{19}$O/$^{18}$O. See text for details.
}
\end{figure}

Presented in Fig.~\ref{fig:fig1}a are the fusion excitation functions 
for $^{16,17,18,19}$O ions 
on a $^{12}$C target. The data for the $^{16,17,18}$O induced reactions have been taken 
from the literature \cite{Kovar79, Eyal76, Heusch82} and in the case of 
$^{18}$O supplemented by our prior measurement \cite{Steinbach14a}. All the 
excitation functions shown are relatively flat at higher energies and at lower 
energies fall exponentially with decreasing energy signaling a barrier-driven 
process. In addition to the overall characteristic shape, one observes for $^{16}$O + $^{12}$C 
a clear increase in the cross-section in the interval 
12 MeV $<$ E$_{c.m.}$ $<$ 18 MeV. Precision measurements in this energy regime 
associate this increased cross-section with well known resonances 
\cite{Frawley82}. Resonances in the case of the other oxygen isotopes are not 
as discernible.

In addition to the fusion excitation function for the $\beta$-stable nuclei, 
the fusion excitation function for $^{19}$O + $^{12}$C is also shown in Fig.~\ref{fig:fig1}a.
Although these cross-sections had been previously reported \cite{Singh17} an improved calibration
has allowed a two to four-fold
improvement in the statistical quality of the data presented \cite{Vadasthesis18}. 
Identification of heavy fusion products was accomplished through use of an energy/time-of-flight approach \cite{Steinbach14a}
with the heavy products detected in annular, double-sided silicon detectors. 
The new calibration of the silicon detectors  allowed inclusion of heavy fusion products over a wider azimuthal range than previously reported.
In these detectors, for each incident particle,
electrons and holes were independently collected. To ensure correct identification of a heavy fusion product, it was 
required that the energy associated with both holes and electrons agree to within 
$\pm$0.5 MeV. The improved energy calibration has resulted in 
the reduction of the number of particles rejected by this energy constraint. 
The re-analyzed cross-section for $^{18}$O + $^{12}$C measured simultaneously 
agrees with the published cross-sections \cite{Singh17}
within the statistical uncertainties.

The original and the revised cross-sections 
are presented as the open red circles and closed red circles respectively. The revised cross-sections are approximately 14\% larger than those previously published \cite{Singh17}. A few points are noteworthy about the new results. 
With the exception of the lowest energy, the revised cross-sections 
are consistently larger than those originally reported although the new cross-sections typically lie within the uncertainties previously reported. The 
reduction in the uncertainty of the
fusion cross-section with the improved statistics is significant.
For the the lowest energy point the cross-section decreases from 158~mb to 128~mb.
Inspection of the cross-section in Fig.~\ref{fig:fig1}a reveals that typical 
cross-section above the barrier for $^{19}$O is $\sim$1200 mb whereas for
$^{18}$O and $^{16}$O it is $\sim$ 1000 mb. It should be emphasized that the magnitude of this increase 
in the cross-section is robustly measured as a simultaneous measurement 
of $^{18,19}$O + $^{12}$C was performed \cite{Singh17}.

The impact of changes in the nuclear size
on the fusion cross-section, 
absent any dynamics, was assessed by employing the S\~{a}o Paulo model~\cite{Gasques04} to predict the fusion cross-section.
The frozen density distributions were determined using a 
relativistic mean field (RMF) approach using the FSUGOLD interaction~\cite{Ring96,Serot86}.
The results of these calculations, designated RMF-SP, are presented in 
Fig.~\ref{fig:fig1} as solid lines. 
While the predictions of the RMF-SP model is in reasonable agreement with the 
experimental cross-sections for $^{16}$O and $^{18}$O (and the limited existing data for $^{17}$O), in the case of $^{19}$O the RMF-SP model 
systematically underpredicts the measured cross-sections.

We have also compared the measured excitation functions with the predictions of an analytical model based on a parameterization of the Sao Paulo potential
model coupled with a barrier penetration fomalism \cite{Afanasjev12, Beard10, Yakolev10}. This model which has parameterized a large number of reactions is a useful tool for network simulations.
The above-barrier cross-sections predicted by this model are depicted in Fig.~\ref{fig:fig1}a as the dashed lines. The cross-sections from the analytic model lie close to the RMF-SP values but are slightly 
smaller in magnitude. While providing a reasonable description of the $^{16,17,18}$O data, the analytical fusion model also underpredicts the $^{19}$O excitation function.

To examine this enhancement of the fusion cross-section for $^{19}$O more quantitatively and with less sensitivity to systematic uncertainties,  
the ratio of $\sigma(^{19}O)$/$\sigma(^{18}O)$ is also
presented in Fig.~\ref{fig:fig1}b.
To construct this ratio the experimental excitation functions have been fit with a simple barrier penetration formalism\cite{Wong73,Singh17}. 
In the above-barrier regime the ratio is relatively flat with a value of approximately 1.22. The uncertainty in the ratio is depicted by the shaded region.

\begin{figure}
\includegraphics[scale=0.42]{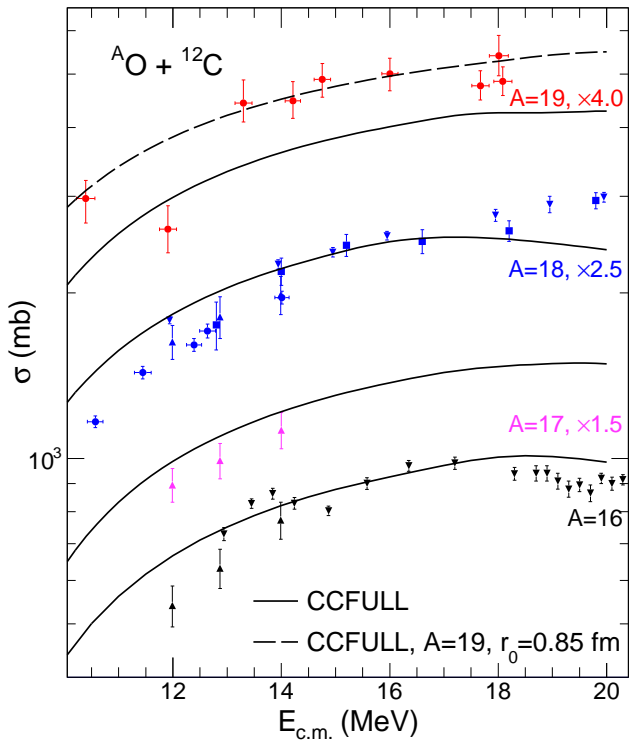}
\caption{\label{fig:fig3} Comparison of CCFULL calculations (solid lines) 
with the experimental excitation functions. 
The long dashed line 
depicts the CCFULL calculations using a radius parameter of
r$_0$=0.85 fm for $^{19}$O. Both data and calculations are scaled by the factors indicated for clarity. 
}
\end{figure}

\begin{table}[h]
\centering
\caption{\label{tab:5/tc}Woods-Saxon parameters used in the CCFULL calculations for the different 
systems.}
\begin{ruledtabular}
\begin{tabular}{c c c c}
 & V$_C$ (MeV) & r$_0$ (fm) & a (fm)\\
\hline
\rule{0pt}{3ex}                                                                
$^{16}$O + $^{12}$C & -256.0 & 0.770 & 0.720 \\
$^{17}$O + $^{12}$C & -256.0 & 0.770 & 0.720 \\
$^{18}$O + $^{12}$C & -281.0 & 0.741 & 0.759 \\
$^{19}$O + $^{12}$C & -300.0 & 0.740 & 0.759 \\
\end{tabular}
\end{ruledtabular}
\end{table}

Given the under-prediction of the measured cross-section 
by the static RMF-SP calculation for  
$^{19}$O , the role of dynamics was considered.
A coupled channels approach \cite{Dasso83} was utilized
to assess the impact of dynamics on the fusion cross-section. 
Coupled channels calculations consider fusion dynamics as the coupling to 
excited states of the colliding nuclei through mutual Coulomb excitation. 
Calculations with this 
coupled-channel approach
have been successful at describing the fusion of stable and
near $\beta$-stable nuclei \cite{Montagnoli13}.
Calculations were performed with the coupled channels code 
CCFULL ~\cite{Hagino99} using the 
Woods-Saxon potential parameters listed in Table~\ref{tab:5/tc}. The value for these parameters was 
chosen consistent 
with the Akyuz-Winther optical model and based on elastic 
scattering data \cite{Akyuz79}. For $^{12}$C, vibrational coupling to 
the 2$^+$ state at 4.439 MeV was included. In the case of $^{16}$O and $^{18}$O, vibrational 
coupling to the 2$^+$ state at 6.917 MeV and 1.982 MeV was accounted 
for while for $^{17}$O and $^{19}$O,
rotational coupling to the excited state 
at 8.466 MeV and 2.779 MeV respectively, a {$\frac{7}{2}^+$} state, was included. 
The resulting fusion 
excitation functions for the various oxygen isotopes are shown in 
Fig.~\ref{fig:fig3} as solid lines. The overall description of the measured fusion excitation 
functions is reasonable with the exception of $^{19}$O. In this case, in order to obtain a 
reasonable description of the excitation function it was necessary to increase the radius 
parameter in 
the Woods-Saxon potential from 0.74 fm to 0.85 fm. The result of the 
latter calculation is represented by the long dashed line. 
The need to arbitrarily increase the radius 
parameter by 15$\%$ in order to reproduce the measured cross-sections could 
be interpreted as the contribution of additional excited states to the 
fusion cross-section.

\begin{figure}
\includegraphics[scale=0.42]{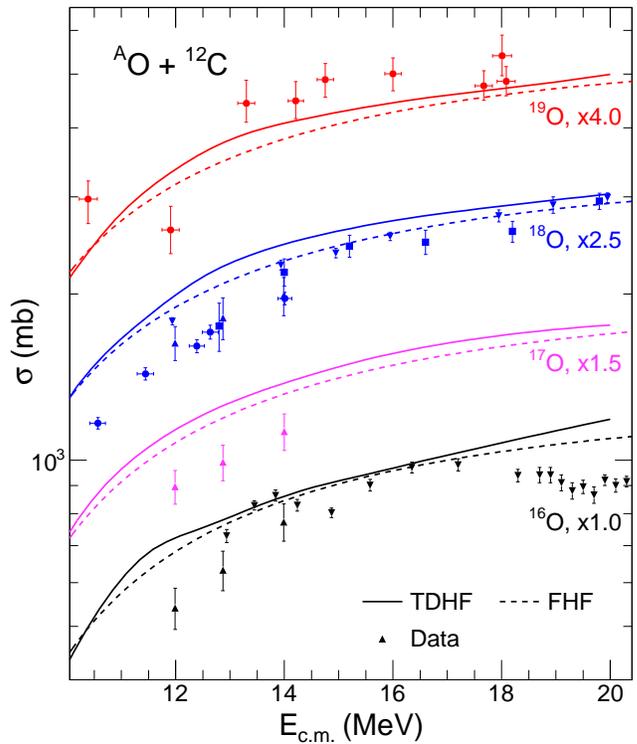}
\caption{\label{fig:fig4} 
Comparison of TDHF and FHF calculations with the experimental excitation functions. 
Both data and calculations are scaled by the factors indicated for clarity. 
}
\end{figure}

To provide an independent 
assessment of the role of dynamics while avoiding the arbitrary truncation at a 
finite number of excited states, we chose to compare the experimental 
data with the predictions of 
a time-dependent Hartree-Fock (TDHF) model.
On general grounds a TDHF approach is well-suited to describing the 
large-amplitude collective motion associated with fusion. 
Recently, advances in theoretical and computational techniques 
allow TDHF calculations to be performed on a 3D Cartesian grid thus eliminating 
artificial symmetry restrictions~\cite{Umar06a}. 
While in the sub-barrier regime in order to accurately describe the 
fusion cross-sections it is necessary to perform density 
constrained TDHF (DC-TDHF) calculations~\cite{Umar12, deSouza13, Steinbach14a}
to
obtain the heavy-ion potentials \cite{Umar06d},
at the above-barrier energies
considered in this work direct TDHF calculations can be performed by initiating collisions
for increasing impact parameters until the maximum impact parameter for fusion is reached.
In practice this was done with an impact parameter precision of 0.02~fm. For odd nuclei
frozen pairing approximation was used in TDHF calculations.

A self-consistent reference point for the TDHF calculations is provided by first
calculating the cross-section by folding frozen Hartree-Fock density 
distributions (FHF). 
As they lack the 
non-local character present in the RMF, these density 
distributions are slightly different from the density distributions 
used in the RMF-SP calculation. Consequently, the fusion cross-sections predicted by the FHF model are slightly larger than those predicted by the RMF-SP.
The predictions of the FHF model (dashed line) are compared with the experimental data in 
Fig.~\ref{fig:fig4}. In the case of $^{16,17,18}$O the predicted fusion 
cross-section matches 
or slightly exceeds the experimental cross-sections. For $^{19}$O however, the predicted cross-sections underpredict the experimental values. 
The impact of the dynamics on the 
cross-section is indicated by the difference between the FHF calculations and the TDHF 
results (solid line). While the impact of the dynamics for these systems is not large, as expected, inclusion of dynamics acts to increase the predicted 
cross-sections. Even with this inclusion of dynamics, however, 
the measured cross-sections 
for fusion of $^{19}$O + $^{12}$C are underpredicted.
This is particularly surprising since for light systems, due to the absence of breakup, the TDHF usually
overpredicts the fusion cross-sections. The relatively small difference between the FHF and TDHF calculations suggests that normally considered excitations of the entrance channel nuclei are unlikely to play a significant role at above barrier energies for light nuclei. This result also suggests that the enhancement in the fusion cross-section for $^{19}$O is {\em not} a standard excitation.

\begin{figure}
\includegraphics[scale=0.42]{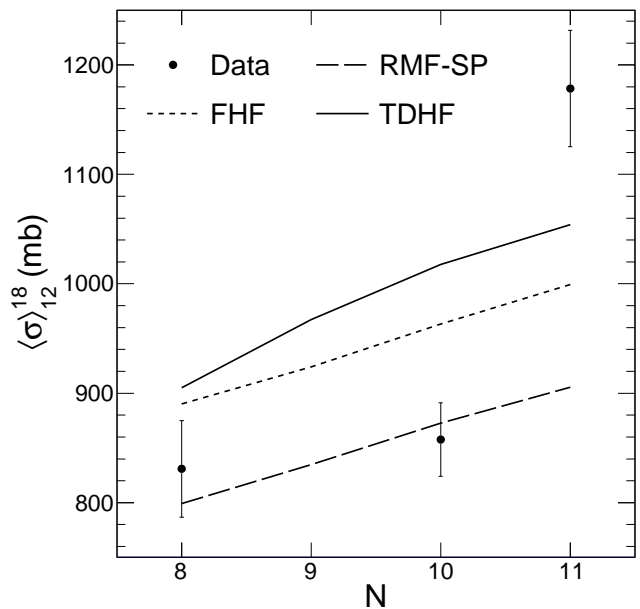}
\caption{\label{fig:fig5} 
Dependence of the average fusion cross-section in the interval
12 MeV $<$ E$_{c.m.}$ $<$ 18 MeV on neutron number. Experimental cross-sections are compared 
with the results of calculations with the RMF-SP, FHF, and TDHF models.
}
\end{figure}

The dependence of the average fusion cross-section on neutron number is examined in 
Fig.~\ref{fig:fig5}. 
The quantity $<$$\sigma$$>$$_{12}^{18}$ 
represents the average fusion cross-section determined over the range 12 MeV $<$ E$_{c.m.}$ $<$ 18 MeV. 
Error bars indicated represent the deviation of the measured cross-section from the average,
energy dependent cross-section. The average energy dependent cross-section was evaluated by 
parameterizing the experimental data \cite{Wong73}. 
While the average cross-section for $^{16}$O and $^{18}$O is comparable, 831 mb and 858 mb, 
respectively, $^{19}$O manifests a significantly higher cross-section of 1178 mb. This 
corresponds to a $\sim$37\% increase in cross-section with the addition of 
a single neutron to $^{18}$O. This $\sim$37\% increase in the 
experimental above-barrier fusion cross-section
with the addition of a single neutron is remarkable. 
It should be 
compared to the increase of the total cross-section in the case of the 
Li isotopes \cite{Blank92}. For the halo nucleus $^{11}$Li, the total reaction 
cross-section increases by 30\% as compared to $^9$Li, a comparable increase. 
This increase in the cross-section for the addition of two neutrons 
is associated with an increase in the 
interaction radius from $\sim$2.4 fm ($^9$Li) to $\sim$3.2 
fm ($^{11}$Li) \cite{Tanihata85b}. It should be appreciated however 
that the high-energy measurement of the total reaction cross-section probes 
effectively the static size of the nuclei. In contrast, the near-barrier fusion 
cross-section 
presented in this work reflects both the static and dynamic 
contributions to the size of the fusing system. The measured interaction cross-section for $^{18,19}$O on
a carbon target at $\sim$900 A.MeV \cite{Ozawa01} exhibits only a modest increase from 1032 $\pm$ 26 mb to 1066 $\pm$ 9 mb with 
increasing neutron number. This result directly indicates that the measured increase in the near-barrier fusion cross-section is not the result 
of simply a larger static size or deformation but results from the low-energy fusion dynamics.
Although the quantity $<$$\sigma$$>$$_{12}^{18}$ cannot be calculated for $^{17}$O due to the lack of experimental data, it can be estimated from Fig.~\ref{fig:fig1}a that if the data followed the smooth behavior predicted 
by the RMF-SP model, one would expect $<$$\sigma$$>$$_{12}^{18}$ cannot be calculated for $^{17}$O due to the lack of experimental data, it can be estimated from Fig$<$$\sigma$$>$$_{12}^{18}$ cannot be calculated for $^{17}$O due to the lack of experimental data, it can be estimated from Fig.~\ref{fig:fig1}a that $<$$\sigma$$>$$_{12}^{18}$
would have a value of $\sim$820 mb. This suggests that the fusion enhancement for $^{19}$O is not due to an odd even effect. This expectation requires experimental verification.

The impact of breakup on the fusion cross-section is a natural question 
to consider as the projectile nuclei become more weakly bound with 
increasing neutron number. 
In the case of $^{11}$Li + $^{208}$Pb, the fusion cross-section predicted by the 
TDHF model exceeded the measured cross-section \cite{Loveland18}. Other models 
provided a similar overprediction \cite{Loveland18}. This overprediction was attributed to the absence of breakup processes in the TDHF model prior 
to fusion.
Projectile breakup is induced either by the target's Coulomb field or 
it's nuclear field.
For Coulomb-driven breakup the breakup probability is expected to scale with atomic number \cite{Paes12}. 
Given the low atomic number of the target nucleus in these reactions it is expected that any breakup 
is dominantly nuclear \cite{Nakamura94}. 
Given the 
range of the nuclear force, this breakup occurs when the two nuclei are close, leading to 
a higher probability of capture. 
Consideration of the Q-values for the nuclei under consideration argues 
against
a dramatic change in the role of breakup for $^{19}$O. 
The Q-value of single neutron removal for $^{19}$O $\rightarrow$ $^{18}$O + n is -3.96 MeV, 
comparable to is -4.14 MeV for $^{17}$O $\rightarrow$ $^{16}$O + n. 
Two-neutron removal for 
$^{19}$O $\rightarrow$ $^{17}$O + 2n is -12.00 MeV, very comparable to the Q$_{2n}$=-12.19 MeV for $^{18}$O. Thus, based on the Q-values a
significant change in the role of capture following neutron breakup is not expected.

The predictions of the average cross-section for the RMF-SP, FHF, and TDHF models are indicated by the dashed, dotted, and solid lines 
respectively. The RMF-SP and FHF exhibit a similar slope with the RMF-SP systematically predicting a smaller 
cross-section (smaller interaction radius) than the 
FHF model. Although the RMF-SP is in 
better agreement with the experimental data for $^{16,18}$O, neither the FHF nor the RMF-SP provides an 
adequate description of the $^{19}$O cross-section. 
The better agreement between the RMF-SP model and the experimental data for 
$^{16,18}$O may be due to its inclusion of
momentum correlations between the nucleons, 
which presumably 
results in more accurate density distributions than the FHF model. From the comparison of the RMF-SP with the 
$^{19}$O cross-section it is clear that $^{19}$O nucleus does not follow the simple scaling 
due to nuclear size as incorporated into the RMF model.
The TDHF model which includes dynamics predicts a 
slightly larger cross-section than the FHF model with a slightly stronger dependence on neutron number. 
However, the TDHF model still  
fails to describe the measured cross-section for
$^{19}$O. The difference between the TDHF 
and FHF calculations indicates the degree to which dynamics, {\em as realized in the TDHF model},
impacts the fusion cross-section at above-barrier 
energies.

Comparison of the experimental data with the static RMF-SP and FHF models 
demonstrates that while the cross-sections for $^{16}$O and $^{18}$O can be 
understood as primarily due to the static size of the nuclei, the enhancement 
of the fusion cross-section for $^{19}$O significantly exceeds the prediction of 
these static models. Even inclusion of dynamics with the TDHF model is insufficient
to explain the increase. This underprediction of experimental cross-sections by 
TDHF also implies that substantially expanding the number of excited 
states included in the 
CCFULL model would also underpredict the experimental data.

The underprediction of the experimental fusion cross-sections by the TDHF model 
suggests 
that extremely large-amplitude collective motion are not sufficiently represented in the TDHF model. 
Though they may be rare, these extremely large-amplitude 
deformations may play a crucial role in the fusion process. 
Whether these large-amplitude deformations are coupled to 
bound states or the continuum is presently unclear. What is clear is that 
high-quality
measurement of above-barrier fusion for an isotopic chain of increasingly neutron-rich nuclei provides an effective means 
to probe these deformations, underscoring the importance of 
such measurements for the most neutron-rich light nuclei accessible.

 \begin{acknowledgments}
This work was supported by the U.S. Department of Energy under Grant No. 
DE-FG02-88ER-40404 (Indiana University), DE-SC0013847 (Vanderbilt University).
CJH is supported in part by U.S. DOE grants DE-FG02-87ER40365 and DE-SC0018083.
ZL gratefully acknowledges support from National Science Foundation under PHY-1613708 (Arizona State University).
 \end{acknowledgments}


\begin{thebibliography}{40}%
\makeatletter
\providecommand \@ifxundefined [1]{%
 \@ifx{#1\undefined}
}%
\providecommand \@ifnum [1]{%
 \ifnum #1\expandafter \@firstoftwo
 \else \expandafter \@secondoftwo
 \fi
}%
\providecommand \@ifx [1]{%
 \ifx #1\expandafter \@firstoftwo
 \else \expandafter \@secondoftwo
 \fi
}%
\providecommand \natexlab [1]{#1}%
\providecommand \enquote  [1]{``#1''}%
\providecommand \bibnamefont  [1]{#1}%
\providecommand \bibfnamefont [1]{#1}%
\providecommand \citenamefont [1]{#1}%
\providecommand \href@noop [0]{\@secondoftwo}%
\providecommand \href [0]{\begingroup \@sanitize@url \@href}%
\providecommand \@href[1]{\@@startlink{#1}\@@href}%
\providecommand \@@href[1]{\endgroup#1\@@endlink}%
\providecommand \@sanitize@url [0]{\catcode `\\12\catcode `\$12\catcode
  `\&12\catcode `\#12\catcode `\^12\catcode `\_12\catcode `\%12\relax}%
\providecommand \@@startlink[1]{}%
\providecommand \@@endlink[0]{}%
\providecommand \url  [0]{\begingroup\@sanitize@url \@url }%
\providecommand \@url [1]{\endgroup\@href {#1}{\urlprefix }}%
\providecommand \urlprefix  [0]{URL }%
\providecommand \Eprint [0]{\href }%
\providecommand \doibase [0]{http://dx.doi.org/}%
\providecommand \selectlanguage [0]{\@gobble}%
\providecommand \bibinfo  [0]{\@secondoftwo}%
\providecommand \bibfield  [0]{\@secondoftwo}%
\providecommand \translation [1]{[#1]}%
\providecommand \BibitemOpen [0]{}%
\providecommand \bibitemStop [0]{}%
\providecommand \bibitemNoStop [0]{.\EOS\space}%
\providecommand \EOS [0]{\spacefactor3000\relax}%
\providecommand \BibitemShut  [1]{\csname bibitem#1\endcsname}%
\let\auto@bib@innerbib\@empty
\bibitem [{\citenamefont {Tanihata}\ \emph
  {et~al.}(1985{\natexlab{a}})\citenamefont {Tanihata} \emph
  {et~al.}}]{Tanihata85b}%
  \BibitemOpen
  \bibfield  {author} {\bibinfo {author} {\bibfnamefont {I.}~\bibnamefont
  {Tanihata}} \emph {et~al.},\ }\href {\doibase 10.1103/physrevlett.55.2676}
  {\bibfield  {journal} {\bibinfo  {journal} {Phys. Rev. Lett.}\ }\textbf
  {\bibinfo {volume} {55}},\ \bibinfo {pages} {2676} (\bibinfo {year}
  {1985}{\natexlab{a}})}\BibitemShut {NoStop}%
\bibitem [{\citenamefont {Tanihata}\ \emph
  {et~al.}(1985{\natexlab{b}})\citenamefont {Tanihata} \emph
  {et~al.}}]{Tanihata85a}%
  \BibitemOpen
  \bibfield  {author} {\bibinfo {author} {\bibfnamefont {I.}~\bibnamefont
  {Tanihata}} \emph {et~al.},\ }\href {\doibase 10.1016/0370-2693(85)90005-X}
  {\bibfield  {journal} {\bibinfo  {journal} {Phys. Lett. B}\ }\textbf
  {\bibinfo {volume} {160}},\ \bibinfo {pages} {380} (\bibinfo {year}
  {1985}{\natexlab{b}})}\BibitemShut {NoStop}%
\bibitem [{\citenamefont {Estrad\'e}\ \emph {et~al.}(2014)\citenamefont
  {Estrad\'e} \emph {et~al.}}]{Estrade14}%
  \BibitemOpen
  \bibfield  {author} {\bibinfo {author} {\bibfnamefont {A.}~\bibnamefont
  {Estrad\'e}} \emph {et~al.},\ }\href {\doibase
  10.1103/PhysRevLett.113.132501} {\bibfield  {journal} {\bibinfo  {journal}
  {Phys. Rev. Lett.}\ }\textbf {\bibinfo {volume} {113}},\ \bibinfo {pages}
  {132501} (\bibinfo {year} {2014})}\BibitemShut {NoStop}%
\bibitem [{\citenamefont {Bagchi}\ \emph {et~al.}(2019)\citenamefont {Bagchi}
  \emph {et~al.}}]{Bagchi19}%
  \BibitemOpen
  \bibfield  {author} {\bibinfo {author} {\bibfnamefont {S.}~\bibnamefont
  {Bagchi}} \emph {et~al.},\ }\href {\doibase 10.1016/j.physletb.2019.01.024}
  {\bibfield  {journal} {\bibinfo  {journal} {Phys. Lett. B}\ }\textbf
  {\bibinfo {volume} {790}},\ \bibinfo {pages} {251 } (\bibinfo {year}
  {2019})}\BibitemShut {NoStop}%
\bibitem [{\citenamefont {Blank}\ \emph {et~al.}(1993)\citenamefont {Blank}
  \emph {et~al.}}]{Blank93}%
  \BibitemOpen
  \bibfield  {author} {\bibinfo {author} {\bibfnamefont {B.}~\bibnamefont
  {Blank}} \emph {et~al.},\ }\href {\doibase 10.1016/0375-9474(93)90294-8}
  {\bibfield  {journal} {\bibinfo  {journal} {Nucl. Phys. A}\ }\textbf
  {\bibinfo {volume} {555}},\ \bibinfo {pages} {408} (\bibinfo {year}
  {1993})}\BibitemShut {NoStop}%
\bibitem [{\citenamefont {Kajino}\ \emph {et~al.}(2019)\citenamefont {Kajino},
  \citenamefont {Aoki}, \citenamefont {Balantekin}, \citenamefont {Diehl},
  \citenamefont {Famiano},\ and\ \citenamefont {Mathews}}]{kajino2019}%
  \BibitemOpen
  \bibfield  {author} {\bibinfo {author} {\bibfnamefont {T.}~\bibnamefont
  {Kajino}}, \bibinfo {author} {\bibfnamefont {W.}~\bibnamefont {Aoki}},
  \bibinfo {author} {\bibfnamefont {A.~B.}\ \bibnamefont {Balantekin}},
  \bibinfo {author} {\bibfnamefont {R.}~\bibnamefont {Diehl}}, \bibinfo
  {author} {\bibfnamefont {M.~A.}\ \bibnamefont {Famiano}}, \ and\ \bibinfo
  {author} {\bibfnamefont {G.~J.}\ \bibnamefont {Mathews}},\ }\href {\doibase
  10.1016/j.ppnp.2019.02.008} {\bibfield  {journal} {\bibinfo  {journal} {Prog.
  Part. Nucl. Phys.}\ }\textbf {\bibinfo {volume} {107}},\ \bibinfo {pages}
  {109} (\bibinfo {year} {2019})}\BibitemShut {NoStop}%
\bibitem [{\citenamefont {{Varinderjit Singh}}\ \emph
  {et~al.}(2017)\citenamefont {{Varinderjit Singh}}, \citenamefont {Vadas},
  \citenamefont {Steinbach}, \citenamefont {Wiggins}, \citenamefont {Hudan},
  \citenamefont {{deSouza}}, \citenamefont {{Zidu Lin}}, \citenamefont
  {Horowitz}, \citenamefont {Baby}, \citenamefont {Kuvin}, \citenamefont
  {{Vandana Tripathi}}, \citenamefont {Wiedenh\"over},\ and\ \citenamefont
  {Umar}}]{Singh17}%
  \BibitemOpen
  \bibfield  {author} {\bibinfo {author} {\bibnamefont {{Varinderjit Singh}}},
  \bibinfo {author} {\bibfnamefont {J.}~\bibnamefont {Vadas}}, \bibinfo
  {author} {\bibfnamefont {T.~K.}\ \bibnamefont {Steinbach}}, \bibinfo {author}
  {\bibfnamefont {B.~B.}\ \bibnamefont {Wiggins}}, \bibinfo {author}
  {\bibfnamefont {S.}~\bibnamefont {Hudan}}, \bibinfo {author} {\bibfnamefont
  {R.~T.}\ \bibnamefont {{deSouza}}}, \bibinfo {author} {\bibnamefont {{Zidu
  Lin}}}, \bibinfo {author} {\bibfnamefont {C.~J.}\ \bibnamefont {Horowitz}},
  \bibinfo {author} {\bibfnamefont {L.~T.}\ \bibnamefont {Baby}}, \bibinfo
  {author} {\bibfnamefont {S.~A.}\ \bibnamefont {Kuvin}}, \bibinfo {author}
  {\bibnamefont {{Vandana Tripathi}}}, \bibinfo {author} {\bibfnamefont
  {I.}~\bibnamefont {Wiedenh\"over}}, \ and\ \bibinfo {author} {\bibfnamefont
  {A.~S.}\ \bibnamefont {Umar}},\ }\href {\doibase
  10.1016/j.physletb.2016.12.017} {\bibfield  {journal} {\bibinfo  {journal}
  {Phys. Lett. B}\ }\textbf {\bibinfo {volume} {765}},\ \bibinfo {pages} {99}
  (\bibinfo {year} {2017})}\BibitemShut {NoStop}%
\bibitem [{\citenamefont {Vadas}\ \emph {et~al.}(2018)\citenamefont {Vadas},
  \citenamefont {Singh}, \citenamefont {Wiggins}, \citenamefont {Huston},
  \citenamefont {Hudan}, \citenamefont {deSouza}, \citenamefont {Lin},
  \citenamefont {Horowitz}, \citenamefont {Chbihi}, \citenamefont {Ackermann},
  \citenamefont {Famiano},\ and\ \citenamefont {Brown}}]{Vadas18}%
  \BibitemOpen
  \bibfield  {author} {\bibinfo {author} {\bibfnamefont {J.}~\bibnamefont
  {Vadas}}, \bibinfo {author} {\bibfnamefont {V.}~\bibnamefont {Singh}},
  \bibinfo {author} {\bibfnamefont {B.~B.}\ \bibnamefont {Wiggins}}, \bibinfo
  {author} {\bibfnamefont {J.}~\bibnamefont {Huston}}, \bibinfo {author}
  {\bibfnamefont {S.}~\bibnamefont {Hudan}}, \bibinfo {author} {\bibfnamefont
  {R.~T.}\ \bibnamefont {deSouza}}, \bibinfo {author} {\bibfnamefont
  {Z.}~\bibnamefont {Lin}}, \bibinfo {author} {\bibfnamefont {C.~J.}\
  \bibnamefont {Horowitz}}, \bibinfo {author} {\bibfnamefont {A.}~\bibnamefont
  {Chbihi}}, \bibinfo {author} {\bibfnamefont {D.}~\bibnamefont {Ackermann}},
  \bibinfo {author} {\bibfnamefont {M.}~\bibnamefont {Famiano}}, \ and\
  \bibinfo {author} {\bibfnamefont {K.~W.}\ \bibnamefont {Brown}},\ }\href
  {\doibase 10.1103/PhysRevC.97.031601} {\bibfield  {journal} {\bibinfo
  {journal} {Phys. Rev. C}\ }\textbf {\bibinfo {volume} {97}},\ \bibinfo
  {pages} {031601(R)} (\bibinfo {year} {2018})}\BibitemShut {NoStop}%
\bibitem [{\citenamefont {Horowitz}\ \emph {et~al.}(2008)\citenamefont
  {Horowitz}, \citenamefont {Dussan},\ and\ \citenamefont
  {Berry}}]{Horowitz08}%
  \BibitemOpen
  \bibfield  {author} {\bibinfo {author} {\bibfnamefont {C.~J.}\ \bibnamefont
  {Horowitz}}, \bibinfo {author} {\bibfnamefont {H.}~\bibnamefont {Dussan}}, \
  and\ \bibinfo {author} {\bibfnamefont {D.~K.}\ \bibnamefont {Berry}},\ }\href
  {\doibase 10.1103/PhysRevC.77.045807} {\bibfield  {journal} {\bibinfo
  {journal} {Phys. Rev. C}\ }\textbf {\bibinfo {volume} {77}},\ \bibinfo
  {pages} {045807} (\bibinfo {year} {2008})}\BibitemShut {NoStop}%
\bibitem [{\citenamefont {RIKEN}()}]{RIKEN}%
  \BibitemOpen
  \bibfield  {author} {\bibinfo {author} {\bibnamefont {RIKEN}},\ }\href
  {http://www.riken.jp/} {}\bibinfo {note} {{N}ishina {C}enter for
  {A}ccelerator-{B}ased {S}cience, {J}apan}\BibitemShut {NoStop}%
\bibitem [{\citenamefont {GANIL}()}]{GANIL}%
  \BibitemOpen
  \bibfield  {author} {\bibinfo {author} {\bibnamefont {GANIL}},\ }\href
  {http://www.ganil-spiral2.eu/} {}\bibinfo {note} {{G}rand {A}ccel\'erateur
  {N}ational d'{I}ons {L}ourds, {C}aen, {F}rance}\BibitemShut {NoStop}%
\bibitem [{\citenamefont {NSCL}()}]{NSCL}%
  \BibitemOpen
  \bibfield  {author} {\bibinfo {author} {\bibnamefont {NSCL}},\ }\href
  {http://www.nscl.msu.edu} {}\bibinfo {note} {{N}ational {S}uperconducting
  {C}yclotron {L}aboratory, {M}ichigan {S}tate {U}niversity, {USA}}\BibitemShut
  {NoStop}%
\bibitem [{\citenamefont {FRIB}()}]{FRIB}%
  \BibitemOpen
  \bibfield  {author} {\bibinfo {author} {\bibnamefont {FRIB}},\ }\href
  {http://frib.msu.edu} {}\bibinfo {note} {{F}acility for {R}are {I}sotope
  {B}eams, {M}ichigan {S}tate {U}niversity, USA}\BibitemShut {NoStop}%
\bibitem [{\citenamefont {Canto}\ \emph {et~al.}(2009)\citenamefont {Canto},
  \citenamefont {Gomes}, \citenamefont {Lubian}, \citenamefont {Chamon},\ and\
  \citenamefont {Crema}}]{Canto09}%
  \BibitemOpen
  \bibfield  {author} {\bibinfo {author} {\bibfnamefont {L.}~\bibnamefont
  {Canto}}, \bibinfo {author} {\bibfnamefont {P.}~\bibnamefont {Gomes}},
  \bibinfo {author} {\bibfnamefont {J.}~\bibnamefont {Lubian}}, \bibinfo
  {author} {\bibfnamefont {L.}~\bibnamefont {Chamon}}, \ and\ \bibinfo {author}
  {\bibfnamefont {E.}~\bibnamefont {Crema}},\ }\href {\doibase
  10.1016/j.nuclphysa.2009.02.001} {\bibfield  {journal} {\bibinfo  {journal}
  {Nucl. Phys. A}\ }\textbf {\bibinfo {volume} {821}},\ \bibinfo {pages} {51}
  (\bibinfo {year} {2009})}\BibitemShut {NoStop}%
\bibitem [{\citenamefont {Afanasjev}\ \emph {et~al.}(2012)\citenamefont
  {Afanasjev}, \citenamefont {Beard}, \citenamefont {Chugunov}, \citenamefont
  {Wiescher},\ and\ \citenamefont {Yakolev}}]{Afanasjev12}%
  \BibitemOpen
  \bibfield  {author} {\bibinfo {author} {\bibfnamefont {A.}~\bibnamefont
  {Afanasjev}}, \bibinfo {author} {\bibfnamefont {M.}~\bibnamefont {Beard}},
  \bibinfo {author} {\bibfnamefont {A.~I.}\ \bibnamefont {Chugunov}}, \bibinfo
  {author} {\bibfnamefont {M.}~\bibnamefont {Wiescher}}, \ and\ \bibinfo
  {author} {\bibfnamefont {D.}~\bibnamefont {Yakolev}},\ }\href {\doibase
  10.1103/PhysRevC.85.054615} {\bibfield  {journal} {\bibinfo  {journal} {Phys.
  Rev. C}\ }\textbf {\bibinfo {volume} {85}},\ \bibinfo {pages} {054615}
  (\bibinfo {year} {2012})}\BibitemShut {NoStop}%
\bibitem [{\citenamefont {Kovar}\ \emph {et~al.}(1979)\citenamefont {Kovar}
  \emph {et~al.}}]{Kovar79}%
  \BibitemOpen
  \bibfield  {author} {\bibinfo {author} {\bibfnamefont {D.~G.}\ \bibnamefont
  {Kovar}} \emph {et~al.},\ }\href {\doibase 10.1103/PhysRevC.20.1305}
  {\bibfield  {journal} {\bibinfo  {journal} {Phys. Rev. C}\ }\textbf {\bibinfo
  {volume} {20}},\ \bibinfo {pages} {1305} (\bibinfo {year}
  {1979})}\BibitemShut {NoStop}%
\bibitem [{\citenamefont {Eyal}\ \emph {et~al.}(1976)\citenamefont {Eyal},
  \citenamefont {Beckerman}, \citenamefont {Chechik}, \citenamefont
  {Fraenkel},\ and\ \citenamefont {Stocker}}]{Eyal76}%
  \BibitemOpen
  \bibfield  {author} {\bibinfo {author} {\bibfnamefont {Y.}~\bibnamefont
  {Eyal}}, \bibinfo {author} {\bibfnamefont {M.}~\bibnamefont {Beckerman}},
  \bibinfo {author} {\bibfnamefont {R.}~\bibnamefont {Chechik}}, \bibinfo
  {author} {\bibfnamefont {Z.}~\bibnamefont {Fraenkel}}, \ and\ \bibinfo
  {author} {\bibfnamefont {H.}~\bibnamefont {Stocker}},\ }\href {\doibase
  10.1103/PhysRevC.13.1527} {\bibfield  {journal} {\bibinfo  {journal} {Phys.
  Rev. C}\ }\textbf {\bibinfo {volume} {13}},\ \bibinfo {pages} {1527}
  (\bibinfo {year} {1976})}\BibitemShut {NoStop}%
\bibitem [{\citenamefont {Heusch}\ \emph {et~al.}(1982)\citenamefont {Heusch},
  \citenamefont {Beck}, \citenamefont {Coffin}, \citenamefont {Engelstein},
  \citenamefont {Freeman}, \citenamefont {Guillaume}, \citenamefont {Haas},\
  and\ \citenamefont {Wagner}}]{Heusch82}%
  \BibitemOpen
  \bibfield  {author} {\bibinfo {author} {\bibfnamefont {B.}~\bibnamefont
  {Heusch}}, \bibinfo {author} {\bibfnamefont {C.}~\bibnamefont {Beck}},
  \bibinfo {author} {\bibfnamefont {J.~P.}\ \bibnamefont {Coffin}}, \bibinfo
  {author} {\bibfnamefont {P.}~\bibnamefont {Engelstein}}, \bibinfo {author}
  {\bibfnamefont {R.~M.}\ \bibnamefont {Freeman}}, \bibinfo {author}
  {\bibfnamefont {G.}~\bibnamefont {Guillaume}}, \bibinfo {author}
  {\bibfnamefont {F.}~\bibnamefont {Haas}}, \ and\ \bibinfo {author}
  {\bibfnamefont {P.}~\bibnamefont {Wagner}},\ }\href {\doibase
  10.1103/PhysRevC.26.542} {\bibfield  {journal} {\bibinfo  {journal} {Phys.
  Rev. C}\ }\textbf {\bibinfo {volume} {26}},\ \bibinfo {pages} {542} (\bibinfo
  {year} {1982})}\BibitemShut {NoStop}%
\bibitem [{\citenamefont {Steinbach}\ \emph {et~al.}(2014)\citenamefont
  {Steinbach}, \citenamefont {Vadas}, \citenamefont {Schmidt}, \citenamefont
  {Haycraft}, \citenamefont {Hudan}, \citenamefont {deSouza}, \citenamefont
  {Baby}, \citenamefont {Kuvin}, \citenamefont {Wiedenh\"over}, \citenamefont
  {Umar},\ and\ \citenamefont {Oberacker}}]{Steinbach14a}%
  \BibitemOpen
  \bibfield  {author} {\bibinfo {author} {\bibfnamefont {T.~K.}\ \bibnamefont
  {Steinbach}}, \bibinfo {author} {\bibfnamefont {J.}~\bibnamefont {Vadas}},
  \bibinfo {author} {\bibfnamefont {J.}~\bibnamefont {Schmidt}}, \bibinfo
  {author} {\bibfnamefont {C.}~\bibnamefont {Haycraft}}, \bibinfo {author}
  {\bibfnamefont {S.}~\bibnamefont {Hudan}}, \bibinfo {author} {\bibfnamefont
  {R.~T.}\ \bibnamefont {deSouza}}, \bibinfo {author} {\bibfnamefont {L.~T.}\
  \bibnamefont {Baby}}, \bibinfo {author} {\bibfnamefont {S.~A.}\ \bibnamefont
  {Kuvin}}, \bibinfo {author} {\bibfnamefont {I.}~\bibnamefont
  {Wiedenh\"over}}, \bibinfo {author} {\bibfnamefont {A.~S.}\ \bibnamefont
  {Umar}}, \ and\ \bibinfo {author} {\bibfnamefont {V.~E.}\ \bibnamefont
  {Oberacker}},\ }\href {\doibase 10.1103/PhysRevC.90.041603} {\bibfield
  {journal} {\bibinfo  {journal} {Phys. Rev. C}\ }\textbf {\bibinfo {volume}
  {90}},\ \bibinfo {pages} {041603(R)} (\bibinfo {year} {2014})}\BibitemShut
  {NoStop}%
\bibitem [{\citenamefont {Frawley}\ \emph {et~al.}(1982)\citenamefont
  {Frawley}, \citenamefont {Fletcher},\ and\ \citenamefont
  {Dennis}}]{Frawley82}%
  \BibitemOpen
  \bibfield  {author} {\bibinfo {author} {\bibfnamefont {A.~D.}\ \bibnamefont
  {Frawley}}, \bibinfo {author} {\bibfnamefont {N.~R.}\ \bibnamefont
  {Fletcher}}, \ and\ \bibinfo {author} {\bibfnamefont {L.~C.}\ \bibnamefont
  {Dennis}},\ }\href {\doibase 10.1103/PhysRevC.25.860} {\bibfield  {journal}
  {\bibinfo  {journal} {Phys. Rev. C}\ }\textbf {\bibinfo {volume} {25}},\
  \bibinfo {pages} {860} (\bibinfo {year} {1982})}\BibitemShut {NoStop}%
\bibitem [{\citenamefont {Vadas}(2018)}]{Vadasthesis18}%
  \BibitemOpen
  \bibfield  {author} {\bibinfo {author} {\bibfnamefont {J.}~\bibnamefont
  {Vadas}},\ }\emph {\bibinfo {title} {{P}robing the {F}usion of
  {N}eutron-{R}ich {N}uclei with {M}odern {R}adioactive {B}eam {F}acilities}},\
  \href@noop {} {Ph.D. thesis},\ \bibinfo  {school} {Indiana University}
  (\bibinfo {year} {2018})\BibitemShut {NoStop}%
\bibitem [{\citenamefont {Gasques}\ \emph {et~al.}(2004)\citenamefont
  {Gasques}, \citenamefont {Chamon}, \citenamefont {Pereira}, \citenamefont
  {Alvarez}, \citenamefont {Rossi}, \citenamefont {Silva},\ and\ \citenamefont
  {Carlson}}]{Gasques04}%
  \BibitemOpen
  \bibfield  {author} {\bibinfo {author} {\bibfnamefont {L.~R.}\ \bibnamefont
  {Gasques}}, \bibinfo {author} {\bibfnamefont {L.~C.}\ \bibnamefont {Chamon}},
  \bibinfo {author} {\bibfnamefont {D.}~\bibnamefont {Pereira}}, \bibinfo
  {author} {\bibfnamefont {M.~A.~G.}\ \bibnamefont {Alvarez}}, \bibinfo
  {author} {\bibfnamefont {E.~S.}\ \bibnamefont {Rossi}}, \bibinfo {author}
  {\bibfnamefont {C.~P.}\ \bibnamefont {Silva}}, \ and\ \bibinfo {author}
  {\bibfnamefont {B.~V.}\ \bibnamefont {Carlson}},\ }\href {\doibase
  10.1103/PhysRevC.69.034603} {\bibfield  {journal} {\bibinfo  {journal} {Phys.
  Rev. C}\ }\textbf {\bibinfo {volume} {69}},\ \bibinfo {pages} {034603}
  (\bibinfo {year} {2004})}\BibitemShut {NoStop}%
\bibitem [{\citenamefont {Ring}(1996)}]{Ring96}%
  \BibitemOpen
  \bibfield  {author} {\bibinfo {author} {\bibfnamefont {P.}~\bibnamefont
  {Ring}},\ }\href {\doibase 10.1016/0146-6410(96)00054-3} {\bibfield
  {journal} {\bibinfo  {journal} {Prog. Part. Nucl. Phys.}\ }\textbf {\bibinfo
  {volume} {37}},\ \bibinfo {pages} {193} (\bibinfo {year} {1996})}\BibitemShut
  {NoStop}%
\bibitem [{\citenamefont {Serot}\ and\ \citenamefont
  {Walecka}(1986)}]{Serot86}%
  \BibitemOpen
  \bibfield  {author} {\bibinfo {author} {\bibfnamefont {B.~D.}\ \bibnamefont
  {Serot}}\ and\ \bibinfo {author} {\bibfnamefont {J.~D.}\ \bibnamefont
  {Walecka}},\ }\href@noop {} {\bibfield  {journal} {\bibinfo  {journal} {Adv.
  Nucl. Phys.}\ }\textbf {\bibinfo {volume} {16}},\ \bibinfo {pages} {1}
  (\bibinfo {year} {1986})}\BibitemShut {NoStop}%
\bibitem [{\citenamefont {Beard}\ \emph {et~al.}(2010)\citenamefont {Beard},
  \citenamefont {Afanasjev}, \citenamefont {Chamon},\ and\ \citenamefont
  {Gasques}}]{Beard10}%
  \BibitemOpen
  \bibfield  {author} {\bibinfo {author} {\bibfnamefont {M.}~\bibnamefont
  {Beard}}, \bibinfo {author} {\bibfnamefont {A.}~\bibnamefont {Afanasjev}},
  \bibinfo {author} {\bibfnamefont {L.}~\bibnamefont {Chamon}}, \ and\ \bibinfo
  {author} {\bibfnamefont {L.}~\bibnamefont {Gasques}},\ }\href {\doibase
  10.1016/j.adt.2010.02.005} {\bibfield  {journal} {\bibinfo  {journal} {Atomic
  Data and Nuclear Data Tables}\ }\textbf {\bibinfo {volume} {96}} (\bibinfo
  {year} {2010}),\ 10.1016/j.adt.2010.02.005}\BibitemShut {NoStop}%
\bibitem [{\citenamefont {Yakolev}\ \emph {et~al.}(2010)\citenamefont
  {Yakolev}, \citenamefont {Beard}, \citenamefont {Gasques},\ and\
  \citenamefont {Wiescher}}]{Yakolev10}%
  \BibitemOpen
  \bibfield  {author} {\bibinfo {author} {\bibfnamefont {D.}~\bibnamefont
  {Yakolev}}, \bibinfo {author} {\bibfnamefont {M.}~\bibnamefont {Beard}},
  \bibinfo {author} {\bibfnamefont {L.}~\bibnamefont {Gasques}}, \ and\
  \bibinfo {author} {\bibfnamefont {M.}~\bibnamefont {Wiescher}},\ }\href
  {\doibase 10.1103/PhysRevC.82.044609} {\bibfield  {journal} {\bibinfo
  {journal} {Phys. Rev. C}\ }\textbf {\bibinfo {volume} {82}},\ \bibinfo
  {pages} {044609} (\bibinfo {year} {2010})}\BibitemShut {NoStop}%
\bibitem [{\citenamefont {Wong}(1973)}]{Wong73}%
  \BibitemOpen
  \bibfield  {author} {\bibinfo {author} {\bibfnamefont {C.~Y.}\ \bibnamefont
  {Wong}},\ }\href {\doibase 10.1103/PhysRevLett.31.766} {\bibfield  {journal}
  {\bibinfo  {journal} {Phys. Rev. Lett.}\ }\textbf {\bibinfo {volume} {31}},\
  \bibinfo {pages} {766} (\bibinfo {year} {1973})}\BibitemShut {NoStop}%
\bibitem [{\citenamefont {Dasso}\ \emph {et~al.}(1983)\citenamefont {Dasso},
  \citenamefont {Landowne},\ and\ \citenamefont {Winther}}]{Dasso83}%
  \BibitemOpen
  \bibfield  {author} {\bibinfo {author} {\bibfnamefont {C.~H.}\ \bibnamefont
  {Dasso}}, \bibinfo {author} {\bibfnamefont {S.}~\bibnamefont {Landowne}}, \
  and\ \bibinfo {author} {\bibfnamefont {A.}~\bibnamefont {Winther}},\ }\href
  {\doibase 10.1016/0375-9474(83)90578-x} {\bibfield  {journal} {\bibinfo
  {journal} {Nucl. Phys. A}\ }\textbf {\bibinfo {volume} {405}},\ \bibinfo
  {pages} {381} (\bibinfo {year} {1983})}\BibitemShut {NoStop}%
\bibitem [{\citenamefont {Montagnoli}\ \emph {et~al.}(2013)\citenamefont
  {Montagnoli} \emph {et~al.}}]{Montagnoli13}%
  \BibitemOpen
  \bibfield  {author} {\bibinfo {author} {\bibfnamefont {G.}~\bibnamefont
  {Montagnoli}} \emph {et~al.},\ }\href {\doibase 10.1103/PhysRevC.87.014611}
  {\bibfield  {journal} {\bibinfo  {journal} {Phys. Rev. C}\ }\textbf {\bibinfo
  {volume} {87}},\ \bibinfo {pages} {014611} (\bibinfo {year}
  {2013})}\BibitemShut {NoStop}%
\bibitem [{\citenamefont {Hagino}\ \emph {et~al.}(1999)\citenamefont {Hagino},
  \citenamefont {Rowley},\ and\ \citenamefont {Kruppa}}]{Hagino99}%
  \BibitemOpen
  \bibfield  {author} {\bibinfo {author} {\bibfnamefont {K.}~\bibnamefont
  {Hagino}}, \bibinfo {author} {\bibfnamefont {N.}~\bibnamefont {Rowley}}, \
  and\ \bibinfo {author} {\bibfnamefont {A.~T.}\ \bibnamefont {Kruppa}},\
  }\href {\doibase 10.1016/s0010-4655(99)00243-x} {\bibfield  {journal}
  {\bibinfo  {journal} {Comput. Phys. Commun.}\ }\textbf {\bibinfo {volume}
  {123}},\ \bibinfo {pages} {143} (\bibinfo {year} {1999})}\BibitemShut
  {NoStop}%
\bibitem [{\citenamefont {Akyuz}\ and\ \citenamefont {Winther}()}]{Akyuz79}%
  \BibitemOpen
  \bibfield  {author} {\bibinfo {author} {\bibfnamefont {O.}~\bibnamefont
  {Akyuz}}\ and\ \bibinfo {author} {\bibfnamefont {A.}~\bibnamefont
  {Winther}},\ }in\ \href@noop {} {\emph {\bibinfo {booktitle} {Proceedings of
  the Enrico Fermi School of Physics 1979}}},\ \bibinfo {editor} {edited by\
  \bibinfo {editor} {\bibfnamefont {R.~A.}\ \bibnamefont {Broglia}}, \bibinfo
  {editor} {\bibfnamefont {C.}~\bibnamefont {.H.Dasso}}, \ and\ \bibinfo
  {editor} {\bibfnamefont {R.}~\bibnamefont {Ricci}}}\ (\bibinfo  {publisher}
  {North-Holland})\BibitemShut {NoStop}%
\bibitem [{\citenamefont {Umar}\ and\ \citenamefont
  {Oberacker}(2006{\natexlab{a}})}]{Umar06a}%
  \BibitemOpen
  \bibfield  {author} {\bibinfo {author} {\bibfnamefont {A.~S.}\ \bibnamefont
  {Umar}}\ and\ \bibinfo {author} {\bibfnamefont {V.~E.}\ \bibnamefont
  {Oberacker}},\ }\href {\doibase 10.1103/PhysRevC.73.054607} {\bibfield
  {journal} {\bibinfo  {journal} {Phys. Rev. C}\ }\textbf {\bibinfo {volume}
  {73}},\ \bibinfo {pages} {054607} (\bibinfo {year}
  {2006}{\natexlab{a}})}\BibitemShut {NoStop}%
\bibitem [{\citenamefont {Umar}\ \emph {et~al.}(2012)\citenamefont {Umar},
  \citenamefont {Oberacker},\ and\ \citenamefont {Horowitz}}]{Umar12}%
  \BibitemOpen
  \bibfield  {author} {\bibinfo {author} {\bibfnamefont {A.~S.}\ \bibnamefont
  {Umar}}, \bibinfo {author} {\bibfnamefont {V.~E.}\ \bibnamefont {Oberacker}},
  \ and\ \bibinfo {author} {\bibfnamefont {C.~J.}\ \bibnamefont {Horowitz}},\
  }\href {\doibase 10.1103/PhysRevC.85.055801} {\bibfield  {journal} {\bibinfo
  {journal} {Phys. Rev. C}\ }\textbf {\bibinfo {volume} {85}},\ \bibinfo
  {pages} {055801} (\bibinfo {year} {2012})}\BibitemShut {NoStop}%
\bibitem [{\citenamefont {deSouza}\ \emph {et~al.}(2013)\citenamefont
  {deSouza}, \citenamefont {Hudan}, \citenamefont {Oberacker},\ and\
  \citenamefont {Umar}}]{deSouza13}%
  \BibitemOpen
  \bibfield  {author} {\bibinfo {author} {\bibfnamefont {R.~T.}\ \bibnamefont
  {deSouza}}, \bibinfo {author} {\bibfnamefont {S.}~\bibnamefont {Hudan}},
  \bibinfo {author} {\bibfnamefont {V.~E.}\ \bibnamefont {Oberacker}}, \ and\
  \bibinfo {author} {\bibfnamefont {A.~S.}\ \bibnamefont {Umar}},\ }\href
  {\doibase 10.1103/PhysRevC.88.014602} {\bibfield  {journal} {\bibinfo
  {journal} {Phys. Rev. C}\ }\textbf {\bibinfo {volume} {88}},\ \bibinfo
  {pages} {014602} (\bibinfo {year} {2013})}\BibitemShut {NoStop}%
\bibitem [{\citenamefont {Umar}\ and\ \citenamefont
  {Oberacker}(2006{\natexlab{b}})}]{Umar06d}%
  \BibitemOpen
  \bibfield  {author} {\bibinfo {author} {\bibfnamefont {A.~S.}\ \bibnamefont
  {Umar}}\ and\ \bibinfo {author} {\bibfnamefont {V.~E.}\ \bibnamefont
  {Oberacker}},\ }\href {\doibase 10.1103/PhysRevC.74.021601} {\bibfield
  {journal} {\bibinfo  {journal} {Phys. Rev. C}\ }\textbf {\bibinfo {volume}
  {74}},\ \bibinfo {pages} {021601(R)} (\bibinfo {year}
  {2006}{\natexlab{b}})}\BibitemShut {NoStop}%
\bibitem [{\citenamefont {Blank}(1992)}]{Blank92}%
  \BibitemOpen
  \bibfield  {author} {\bibinfo {author} {\bibfnamefont {B.}~\bibnamefont
  {Blank}},\ }\href {\doibase 10.1007/BF01289813} {\bibfield  {journal}
  {\bibinfo  {journal} {Z. Phys. A}\ }\textbf {\bibinfo {volume} {343}},\
  \bibinfo {pages} {375} (\bibinfo {year} {1992})}\BibitemShut {NoStop}%
\bibitem [{\citenamefont {Ozawa}\ \emph {et~al.}(2001)\citenamefont {Ozawa},
  \citenamefont {Suzuki},\ and\ \citenamefont {Tanihata}}]{Ozawa01}%
  \BibitemOpen
  \bibfield  {author} {\bibinfo {author} {\bibfnamefont {A.}~\bibnamefont
  {Ozawa}}, \bibinfo {author} {\bibfnamefont {T.}~\bibnamefont {Suzuki}}, \
  and\ \bibinfo {author} {\bibfnamefont {I.}~\bibnamefont {Tanihata}},\
  }\href@noop {} {\bibfield  {journal} {\bibinfo  {journal} {Nucl. Phys. A}\
  }\textbf {\bibinfo {volume} {603}},\ \bibinfo {pages} {32} (\bibinfo {year}
  {2001})}\BibitemShut {NoStop}%
\bibitem [{\citenamefont {Loveland}\ \emph {et~al.}(2018)\citenamefont
  {Loveland}, \citenamefont {Vinodkumar}, \citenamefont {Yanez}, \citenamefont
  {Yao}, \citenamefont {King}, \citenamefont {Lassen},\ and\ \citenamefont
  {Rojas}}]{Loveland18}%
  \BibitemOpen
  \bibfield  {author} {\bibinfo {author} {\bibfnamefont {W.}~\bibnamefont
  {Loveland}}, \bibinfo {author} {\bibfnamefont {A.~M.}\ \bibnamefont
  {Vinodkumar}}, \bibinfo {author} {\bibfnamefont {R.}~\bibnamefont {Yanez}},
  \bibinfo {author} {\bibfnamefont {L.}~\bibnamefont {Yao}}, \bibinfo {author}
  {\bibfnamefont {J.}~\bibnamefont {King}}, \bibinfo {author} {\bibfnamefont
  {J.}~\bibnamefont {Lassen}}, \ and\ \bibinfo {author} {\bibfnamefont
  {A.}~\bibnamefont {Rojas}},\ }\href {\doibase 10.1140/epja/i2018-12572-8}
  {\bibfield  {journal} {\bibinfo  {journal} {Eur. Phys. J. A}\ }\textbf
  {\bibinfo {volume} {54}},\ \bibinfo {pages} {140} (\bibinfo {year}
  {2018})}\BibitemShut {NoStop}%
\bibitem [{\citenamefont {Paes}\ \emph {et~al.}(2012)\citenamefont {Paes},
  \citenamefont {Lubian}, \citenamefont {Gomes},\ and\ \citenamefont
  {Guimar\"{a}es}}]{Paes12}%
  \BibitemOpen
  \bibfield  {author} {\bibinfo {author} {\bibfnamefont {B.}~\bibnamefont
  {Paes}}, \bibinfo {author} {\bibfnamefont {J.}~\bibnamefont {Lubian}},
  \bibinfo {author} {\bibfnamefont {P.~R.~S.}\ \bibnamefont {Gomes}}, \ and\
  \bibinfo {author} {\bibfnamefont {V.}~\bibnamefont {Guimar\"{a}es}},\ }\href
  {\doibase 10.1016/j.nuclphysa.2012.07.011} {\bibfield  {journal} {\bibinfo
  {journal} {Nucl. Phys. A}\ }\textbf {\bibinfo {volume} {890}},\ \bibinfo
  {pages} {1} (\bibinfo {year} {2012})}\BibitemShut {NoStop}%
\bibitem [{\citenamefont {Nakamura}\ \emph {et~al.}(1994)\citenamefont
  {Nakamura} \emph {et~al.}}]{Nakamura94}%
  \BibitemOpen
  \bibfield  {author} {\bibinfo {author} {\bibfnamefont {T.}~\bibnamefont
  {Nakamura}} \emph {et~al.},\ }\href {\doibase 10.1016/0370-2693(94)91055-3}
  {\bibfield  {journal} {\bibinfo  {journal} {Phys. Lett. B}\ }\textbf
  {\bibinfo {volume} {331}},\ \bibinfo {pages} {296 } (\bibinfo {year}
  {1994})}\BibitemShut {NoStop}%
\end{thebibliography}



%

\end{document}